\documentclass[12pt,onecolumn,journal]{IEEEtran}

\usepackage{amsmath}
\usepackage{amssymb}
\usepackage[dvips]{graphicx}
\usepackage{setspace}
\usepackage{epsfig}
\usepackage{color}

\newcommand{\tSNR}{\text{SNR}}

\newcommand{\figsize}{0.5}

\newcommand{\br}{\mathbf{r}}

\newcommand{\q}{\mathbf{q}}

\newtheorem{theo}{Theorem}

\begin{document}

\title{Performance Analysis of Cognitive Radio Systems with Imperfect Channel Sensing and Estimation}

\author{\authorblockN{Sami Akin and M. Cenk Gursoy}\\
\thanks{S. Akin is with the Institute of Communications Technology, Leibniz University of Hanover, Hanover, Germany 30167,
(E-mail: sami.akin@ikt.uni-hannover.de).}
\thanks{M. Cenk Gursoy is with the Department of Electrical
Engineering and Computer Science, Syracuse University, Syracuse, NY, 13244
(e-mail: mcgursoy@syr.edu).}
\thanks{This work was partially supported by the European Research Council under
Starting Grant--306644.}}

\date{}

\maketitle

\begin{spacing}{1.7}
\begin{abstract}

In cognitive radio systems, employing sensing-based spectrum access strategies, secondary users are required to perform channel sensing in order to detect the activities of licensed primary users in a channel, and in realistic scenarios, channel sensing occurs with possible errors due to miss-detections and false alarms. As another challenge, time-varying fading conditions in the channel between the secondary transmitter and the secondary receiver have to be learned via channel estimation. In this paper, performance of causal channel estimation methods in correlated cognitive radio channels under imperfect channel sensing results is analyzed, and achievable rates under both channel and sensing uncertainty are investigated. Initially, cognitive radio channel model with channel sensing error and channel estimation is described. Then, using pilot symbols, minimum mean square error (MMSE) and linear-MMSE (L-MMSE) estimation methods are employed at the secondary receiver to learn the channel fading coefficients. Expressions for the channel estimates and mean-squared errors (MSE) are determined, and their dependencies on channel sensing results, and pilot symbol period and energy are investigated. Since sensing uncertainty leads to uncertainty in the variance of the additive disturbance, channel estimation strategies and performance are interestingly shown to depend on the sensing reliability. It is further shown that the L-MMSE estimation method, which is in general suboptimal, performs very close to MMSE estimation. Furthermore, assuming the channel estimation errors and the interference introduced by the primary users as zero-mean and Gaussian distributed, achievable rate expressions of linear modulation schemes and Gaussian signaling are determined. Subsequently, the training period, and data and pilot symbol energy allocations are jointly optimized to maximize the achievable rates for both signaling schemes.
\end{abstract}


\end{spacing}

\thispagestyle{empty}
\begin{spacing}{1.95}
\section{Introduction}
Cognitive radio communication has emerged as a new paradigm to tackle down the problem of inefficient use of the spectrum, which arises mainly due to rigid rules allowing only the licensed users to have exclusive access to certain frequency bands. In order to bring flexibility to the spectrum usage, three different strategies, namely underlay, overlay, and interweave operations of cognitive radio systems have been proposed (see e.g., \cite{goldsmith} and references therein). In underlay cognitive radio networks, secondary users are allowed to coexist with the primary users as long as they satisfy strict constraints on the interference inflicted on the primary users. On the other hand, in interweave cognitive radio networks, secondary users initially perform channel sensing and then opportunistically access the spectrum holes. These two spectrum access techniques can be combined for improved performance \cite{kang_1}. For instance, secondary users can sense the activities of the primary users in order to transmit with full power over idle-sensed channels and also communicate over a busy-sensed channel while keeping the interference inflicted on the primary users within tolerable limits.

Due to their significance, channel sensing algorithms and spectrum access policies have recently been extensively studied (see e.g., \cite{Zhao}, \cite{tevfik} and references therein) and a considerable number of challenges and design issues from the perspectives of cognitive secondary and/or primary users were addressed. Several sensing methods, such as matched filtering, cyclo-stationary feature detection, and energy detection were investigated, cooperation among secondary users was considered (\cite{cabric}-\cite{chen_add}). Additionally, wideband channel sensing algorithms were studied. For instance, the authors introduced novel wideband spectrum sensing techniques in \cite{zhi} and \cite{pedram}. The authors in \cite{sheu_sheu} studied the transmission collision between the primary and secondary users, and they optimized the transmission time of the secondary users between consecutive sensing phases to maximize the their throughput. Finally, in \cite{tandra} the authors analyzed the limits of channel sensing methods by analyzing the SNR-wall below which sensing performance will not improve further.

As another challenge for cognitive radios, wireless transmission medium is subject to variations over time due to mobility and/or changing environment. Most practical wireless systems attempt to learn the channel conditions, and due to this, different channel estimation methods and their performances have been investigated in detail in the literature. Transmitting known pilot symbols is generally used as a method to obtain channel state information \cite{pilot}. For instance, the authors in \cite{abou_faycal} focused on pilot symbol assisted modulation approach for transceiver design for time-varying channels, and proposed causal and non-causal estimation algorithms, and compared different algorithms by maximizing the mutual information between input and output over the spacing of pilot symbols. Further results and characterizations on pilot-symbol assisted wireless communications can be found, for instance, in \cite{yunfei}--\cite{taghi}.  Several recent studies have addressed the channel estimation problem in cognitive radio systems as well and focused on pilot allocation strategies. For instance, by minimizing the mean-squared error (MSE) of the least-squares channel estimation method, a practical pilot design method for OFDM-based cognitive radios was proposed in \cite{hu}. The relationship between the optimal pilot pattern and the power loading was investigated in \cite{bojan} under the constraint that primary users did not experience interference above a certain threshold. However, in these studies, channel sensing errors have not been considered.

In a practical cognitive radio setting, channel sensing errors are experienced due to false-alarms and miss-detections, and this sensing uncertainty can have an impact on the design of channel estimation algorithms and pilot placement policies. Additionally, it is of interest to identify how channel sensing errors affect the quality of channel estimation. On the other hand, despite the practical significance of these considerations, there has only been limited work focusing on the joint treatment of channel sensing and estimation. In \cite{gao}, Gao \emph{et al.} addressed channel training and estimation in a multiple-antenna cognitive radio setting. In their model, cognitive users initially listen to the primary users' transmission in order to learn the structure of the covariance matrices of the received signals and perform receive and transmit beamforming in their own transmissions. Following this learning phase, the cognitive users enter into a training phase in which pilot signals are sent and the linear minimum mean-square-error (LMMSE) estimation is performed. In \cite{gursoy-gezici}, we studied the interactions between channel sensing and estimation. More specifically, we investigated the structure and performance of different channel estimation schemes in the presence of sensing uncertainty.

While channel sensing and estimation are critical tasks in a practical setting, the ultimate goal of cognitive users is to perform data transmission and reception. With that, capacity and throughput of cognitive radio systems have been investigated in numerous studies. For example, assuming that the channel between the secondary transmitter and the primary receiver is known by the secondary users, the authors in \cite{amir} investigated the capacity gains offered to the secondary users by opportunistically sharing the spectrum when the communication channels vary due to fading. Liang $\textit{et al.}$ in \cite{ying-chang} studied the problem of optimizing the channel sensing duration to maximize the achievable throughput for the secondary network under the constraint that the primary users are sufficiently protected. Furthermore, under transmit power constraints and also interference power constraints for primary user protection, optimal power allocation strategies for the secondary users were derived in \cite{kang}. More recently, the authors in \cite{stotas} focused on the average achievable throughput and spectrum sensing capabilities compared to the conventional opportunistic spectrum access cognitive radio systems. We  note that in the above-mentioned studies, the authors worked on cognitive radio models in which the secondary users have complete channel side information (CSI) between the secondary receiver and the secondary transmitter, and the secondary transmitter and the primary receiver, and did not consider imperfectly known channel conditions. In \cite{musavian} -- \cite{smith}, the authors addressed the impact of imperfect channel knowledge on the capacity of cognitive radio channels in underlay scenarios under interference power limitations. In these studies, imperfect channel knowledge was regarding the channel conditions in the links between the secondary and primary users rather than the link between the secondary users. Moreover, since underlay schemes were considered, channel sensing was not addressed, and explicit channel estimation methods and the interplay between channel sensing and channel estimation were generally not investigated.

In this paper, we consider a cognitive radio model in which the secondary users communicate over randomly-varying fading channels with memory in the presence of channel sensing uncertainty and with imperfect channel knowledge. Throughout the paper, we assume that only statistical knowledge is available a priori, regarding the channel conditions, noise, primary user activity, and primary users' received signals. In this setting, secondary users initially sense the channel using a quadratic detector, and then perform the estimation of their own channel conditions in a correlated fading environment, and finally establish communication with imperfect sensing and estimation results. In both channel estimation and data transmission, secondary users select their power levels depending on the sensing results. For instance, if the channel is sensed as busy, they can operate with lower power than they would otherwise or cease transmission altogether in order to protect the primary users. Under these assumptions, our key contributions in this paper can be summarized as follows:
\begin{enumerate}
\item We unveil the relationship between imperfect channel sensing and imperfect channel estimation in a fading channel with memory. Therefore, unlike in \cite{gursoy-gezici} in which the estimation of a single fading coefficient is considered, we address the estimation of a block of correlated fading coefficients.
\item We concentrate on two different estimation methods, MMSE and L-MMSE, and obtain analytical expressions for the channel estimates and mean-squared error (MSE) values. We identify the impact of channel sensing uncertainty on the channel estimator structure and channel estimation quality.
\item We characterize the achievable rates when linear modulation schemes or Gaussian input signaling are employed at the secondary transmitter with imperfect CSI at the secondary receiver obtained through channel estimation.
\item We determine efficient energy allocation strategies among data symbols and channel training symbols, and efficient training period values via numerical analysis.
\end{enumerate}

The organization of the rest of the paper is as follows. In Section \ref{sec:channel_model_sensing}, we introduce the cognitive radio channel with channel sensing imperfections. In Section \ref{pilot_symbol_modulation}, we discuss different causal channel estimation techniques, and provide expressions for the channel estimates and MSE values. In Section \ref{achievable_rates}, we determine the rates achieved with linear modulation schemes and Gaussian signaling, and address the achievable-rate-maximizing energy allocation among data and pilot symbols, and also of pilot symbol period. We present the numerical results in Section \ref{numerical_results} and conclude in Section \ref{conclusion}.

\section{Channel Model}\label{sec:channel_model_sensing}
\begin{figure}
\begin{center}
\includegraphics[width =0.5\textwidth]{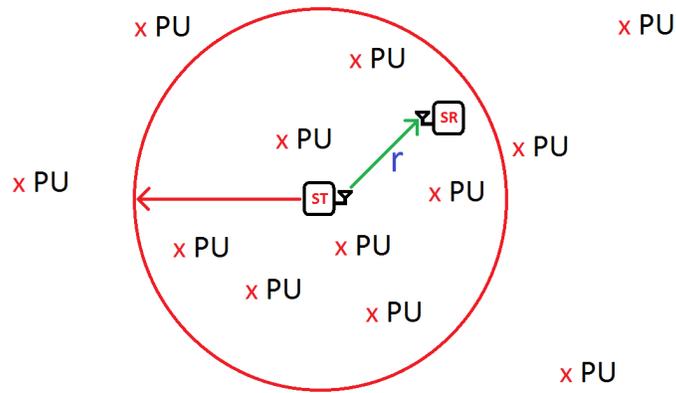}
\caption{Cognitive radio channel model.}\label{fig:res1_ek}
\end{center}
\end{figure}

\begin{figure*}
\begin{center}
\includegraphics[width =1\textwidth]{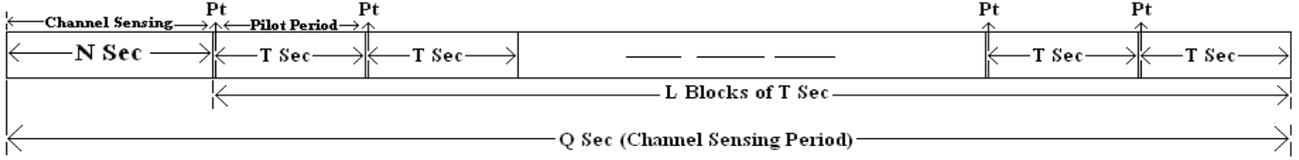}
\vspace{-2cm}
\caption{Frame structure of cognitive radio transmissions.}\label{fig:res1}
\end{center}
\end{figure*}

In this section, we describe a cognitive radio model in which a secondary transmitter and a secondary receiver communicate over a time-selective correlated Rayleigh flat fading channel in the presence of primary users as seen in Figure \ref{fig:res1_ek}. First, the secondary users enter the channel sensing phase and sense the activities of the primary users in the channel in the first $N$ seconds of the frame duration of $Q$ seconds. Based on the sensing results, the power level for the transmission phase is selected. Following channel sensing, the secondary transmitter sends the pilot and data symbols in $L$ blocks of $T$ seconds. This transmission frame structure is depicted in Fig. \ref{fig:res1}. More specifically, in each block of $T$ seconds, the secondary transmitter initially sends a pilot symbol to enable channel estimation at the secondary receiver and subsequently transmits the data. This procedure is repeated  $L$ times until the next channel sensing phase\footnote{Following the channel sensing phase, pilot symbols are inserted periodically, and data symbols are transmitted in between these pilot symbols until the next channel sensing phase.}. Note that sensing is performed with a period of $Q$ seconds and, from the above description, we have $Q = N + LT$. We further assume that the bandwidth available in the system is $B$ and symbol rate is $B$ complex symbols per second.

Meanwhile, the discrete-time channel input-output relation in the $k^{\text{th}}$ symbol period is given by
\begin{align}\label{input-out1}
&y_{k}=r_{k}x_{k}+n_{k}+s_{k}\quad k=1,2,\dots
\end{align}
if the primary users are active in the channel. On the other hand, if there are no active primary users in the channel, we have
\begin{align}\label{input-out2}
&y_{k}=r_{k}x_{k}+n_{k}\quad k=1,2,\dots
\end{align}
In the above equations, $x_{k}$ is the complex-valued channel input and $y_{k}$ is the complex-valued channel output. $\{n_{k}\}$ is assumed to be a sequence of independent and identically distributed (i.i.d.) zero mean Gaussian random variables with variance $\sigma_{n}^{2}$. Furthermore, $r_{k}$ denotes the channel fading coefficient between the secondary transmitter and the secondary receiver, and $\{r_k\}$ is assumed to be a zero-mean Gaussian random process with an auto-correlation function $R_r(\tau)$ and variance $R_r(0)=\sigma_{r}^{2}$. While both the secondary transmitter and the secondary receiver know the channel statistics\footnote{Channel statistics generally vary much more slowly than instantaneous channel realizations and they can be learned via long term observations or estimated by adopting channel models for rural, urban, and suburban environments.}, neither has the prior knowledge of the instantaneous realizations of $\{r_k\}$. Finally, in (\ref{input-out1}), $s_{k}$ represents the sum of the active primary users' \emph{faded} signals arriving at the secondary receiver\footnote{Hence, $s_k$ represents sum of the primary users' signals multiplied with the fading coefficients.}.

We assume that transmissions are subject to average power constraints. In particular, average transmission power in each transmission block of $T$ seconds is $\overline{P}_{0}$ when the channel is sensed as idle, and $\overline{P}_{1}$ when the channel is sensed as busy. Hence, the average transmission energy in each block is $\overline{P}_{0} T$ or $\overline{P}_{1} T$ depending on the sensed primary user activity\footnote{In the subsequent sections, we consider the allocation of this average transmission energy among pilot and data symbols.}. Note that we consider a general scenario in which the secondary users can coexist with the active primary users as long as their transmissions do not result in excessive interference on the primary users. However, if secondary users are not allowed to transmit in a busy-sensed channel, we can set $\overline{P}_{1} = 0$. Hence, the consideration of this two-level transmission power policy gives flexibility and enables better control of the interference inflicted on the primary users. For instance, the power levels $\overline{P}_{0}$ and $\overline{P}_{1}$ can be dictated by average interference power constraints\footnote{For instance, with imperfect sensing, average interference power is \emph{proportional} to $P_d \overline{P}_{1} + (1-P_d) \overline{P}_{0}$ where $P_d$ is the detection probability in channel sensing. This is due to the fact that interference is proportional to $\overline{P}_{1}$ with probability $P_d$ when primary user activity is correctly detected, and is proportional to $\overline{P}_{0}$ with probability $1-P_d$ when primary user activity is missed. Hence, $\overline{P}_{0}$ and $\overline{P}_{1}$ can be selected so that $P_d \overline{P}_{1} + (1-P_d) \overline{P}_{0}$ does not exceed a certain threshold and  average interference power remains within tolerable limits.}.

Finally, we consider a practical scenario in which errors such as miss-detections and false alarms may occur in channel sensing. We denote the correct-detection probability of the active primary users by $P_d$, and the false-alarm probability by $P_f$.

\section{Pilot Symbol Assisted Modulation and Channel Estimation}\label{pilot_symbol_modulation}
In the training phase, periodically inserted pilot symbols, known by both the secondary transmitter and the secondary receiver, are used to estimate the channel fading coefficients. We assume a simple scenario in which the first symbol of each block of $TB$ symbols is transmitted as a pilot symbol every $T$ seconds as seen in Fig. \ref{fig:res1} (i.e., at time indices $N+lT$, $l=0,1,...,L-1$), and the secondary receiver estimates the channel fading coefficients. As described in the channel model, transmission power/energy is assumed to depend on the sensing decision. A similar two-level transmission energy is also considered in the training phase. More specifically, the energy of the pilot symbol is $\mathcal{E}_{t,0}$ when the channel is sensed as idle, and is $\mathcal{E}_{t,1}$ when the channel is sensed as busy. Such a scheme of transmitting at two different energy levels, depending on the sensing results, enables the secondary users to better control the interference inflicted on the primary users. For instance, in order to protect the primary users, interference constraints can be imposed, and the transmission energy/power levels can be determined under such constraints. Further details on interference limitations can be found in \cite{sami2} and \cite{Rui}.

We examine two different channel estimation methods: MMSE and L-MMSE. One pilot symbol, transmitted at the beginning of each block of $TB$ symbols, is used to estimate the channel fading coefficient at the moment the pilot symbol is transmitted and also the following $TB-1$ channel fading coefficients. In other words, the secondary receiver utilizes the past pilot symbol and the correlation structure of the fading to obtain information about the current channel fading coefficient. For the sake of notational simplicity, we henceforth denote $TB$ by $M$, i.e., $M = TB$. We further note that in order for clarity in notation we also assume that the first transmitted symbol after channel sensing, i.e., the first pilot symbol, is transmitted at time 0.

\subsection{MMSE Estimation}
The input-output relation in the training phase is given as
\begin{align}\label{training_one_pilot}
\underbrace{\left[
            \begin{array}{c}
              y_{lM} \\
							y_{(l-1)M}\\
              \vdots \\
              y_{(l-K+1)M} \\
            \end{array}
          \right]}_{Y_{l}}
=\underbrace{\begin{pmatrix}
    0 & 0 &\cdots & 0 & 0 & \sqrt{\mathcal{E}_{t,i}} & 0 & \cdots & 0\\
    0 & 0 &\cdots & 0 & \sqrt{\mathcal{E}_{t,i}} & 0 & 0 & \cdots & 0\\
    \vdots & \vdots & \vdots & \vdots & \vdots & \vdots & \vdots & \cdots & \vdots\\
		0 & \sqrt{\mathcal{E}_{t,i}} & \cdots & 0 & 0 & 0 & 0 & \cdots & 0\\
		\sqrt{\mathcal{E}_{t,i}} & 0 & \cdots & 0 & 0 & 0 & 0 & \cdots & 0
  \end{pmatrix}}_{\q_{l,t,i}}\underbrace{\left[
            \begin{array}{c}
							r_{(l-K+1)M}\\
							r_{(l-K+2)M}\\
							\vdots\\
              r_{lM} \\
							r_{lM+1}\\
              \vdots \\
              r_{(l+1)M-1} \\
            \end{array}
          \right]}_{\br_{l}}&+\underbrace{\left[
            \begin{array}{c}
              w_{lM} \\
							w_{(l-1)M}\\
              \vdots \\
              w_{(l-K+1)M} \\
            \end{array}
          \right]}_{W_{l}}
\end{align}
where $i\in\{0,1\}$, and
\begin{equation*}
W_{l}=\underbrace{[n_{lM},n_{(l-1)M},\cdots,n_{(l-K+1)M}]^{T}}_{N_{l}}
\end{equation*}
or
\begin{equation*}
W_{l}=N_{l}+\underbrace{[s_{lM},s_{(l-1)M},\cdots,s_{(l-K+1)M}]^{T}}_{S_{l}}
\end{equation*}
when the channel is actually idle or busy, respectively\footnote{The superscript $T$ denotes the transpose operation.}. Moreover, $\br_l$ denotes the vector of $M-1$ fading coefficients to be estimated in the $l^{th}$ data transmission block and the last $K$ pilot symbols transmitted before the data symbols to be sent in the $l^{th}$ block, and we define $\q_{l,t,i}$ as the pilot matrix to be used in channel estimation. Here, we assume that the secondary receiver uses the last $K$ transmitted pilot symbols. However, it could be also designed in such a way that the secondary receiver uses all the pilot symbols transmitted after channel sensing. Hence, we further note that the pilot symbol transmitted at time $(l-K+1)M$ is the first pilot symbol transmitted after channel sensing if all pilot symbols transmitted after channel sensing are utilized\footnote{$K$ is a not fixed value for every $l$. For instance, when $l=0$, $K=1$, or when $l=L$, $K=L+1$.}. In other words, the channel estimation method uses the strategy of accumulating past pilot symbol data in order to obtain channel side information regarding the fading coefficients following the last transmitted pilot symbol. Since the channel fading coefficients are correlated and this correlation structure is available at the receiver, the receiver incorporates all available transmitted pilot symbols up to that time in order to estimate the channel fading coefficients. However, we underline that since the correlation among channel fading coefficients is in general decreasing with the increase in the time difference between the channel fading coefficients, it will be sufficient to use a certain number of pilot symbols rather than using all of the transmitted past pilot symbols. Especially, if the channel decorrelates fast, it will be enough to use one pilot symbol to estimate the few number of channel coefficients following the pilot\footnote{In the extreme case in which there is no correlation among the channel fading coefficients at all (i.e., fading is independently changing for every symbol), current estimate by the pilot symbol will be of no use to estimate the forthcoming fading coefficients.}. We refer the interested reader to \cite{pilot} and \cite{abou_faycal}. The receiver obtains the estimate $\widehat{\br}_{0}$ when the channel sensing decision is $\widehat{\mathcal{H}}_{0}$, and obtains $\widehat{\br}_{1}$ when the channel sensing decision is $\widehat{\mathcal{H}}_{1}$. Now, we can express the MMSE estimation as the following optimization problem:
\begin{align}
\min_{\widehat{\br}_{l}}\mathbb{E}\{\|\br_{l}-\widehat{\br}_{l}\|^2\}=&\min_{\widehat{\br}_{l,0},\widehat{\br}_{l,1}}\sum_{j=0}^{1}\Pr\{\widehat{\mathcal{H}}_{j}\}\mathbb{E}\{\|\underbrace{\br_{l}-\widehat{\br}_{l,j}}_{\widetilde{\br}_{l,j}}\|^2|\widehat{\mathcal{H}}_{j}\}\label{mmse_single_pilot}  \\
=&\sum_{j=0}^{1}\Pr\{\widehat{\mathcal{H}}_{j}\}\min_{\widehat{\br}_{l,j}}\mathbb{E}\{\|\widetilde{\br}_{l,j}\|^2|\widehat{\mathcal{H}}_{j}\}\label{mmse_single_pilot2}
\end{align}
where $\Pr\{\widehat{\mathcal{H}}_{0}\}$ and $\Pr\{\widehat{\mathcal{H}}_{1}\}$ are the probabilities of channel being sensed as idle and busy, respectively, and they can be expressed as
\begin{align*}
\Pr\{\widehat{\mathcal{H}}_{0}\}&=\Pr\{\mathcal{H}_{0}\}\Pr\{\widehat{\mathcal{H}}_{0}|\mathcal{H}_{0}\}+\Pr\{\mathcal{H}_{1}\}\Pr\{\widehat{\mathcal{H}}_{0}|\mathcal{H}_{1}\}
=\Pr\{\mathcal{H}_{0}\}(1-P_f)+\Pr\{\mathcal{H}_{1}\}(1-P_d)
\end{align*}
and
\begin{align*}
\Pr\{\widehat{\mathcal{H}}_{1}\}&=\Pr\{\mathcal{H}_{0}\}\Pr\{\widehat{\mathcal{H}}_{1}|\mathcal{H}_{0}\}+\Pr\{\mathcal{H}_{1}\}\Pr\{\widehat{\mathcal{H}}_{1}|\mathcal{H}_{1}\}
=\Pr\{\mathcal{H}_{0}\}P_f+\Pr\{\mathcal{H}_{1}\}P_d
\end{align*}
with $P_f$ and $P_d$ again denoting the false-alarm and detection probabilities, respectively. Note that $\widetilde{\br}_{l,0}$ and $\widetilde{\br}_{l,1}$ are the channel estimation error vectors, and $\sigma_{\widetilde{r}_{l,0,k}}^2$ and $\sigma_{\widetilde{r}_{l,1,k}}^2$ represent the variances of the $k^{\text{th}}$ elements of $\widetilde{\br}_{l,0}$ and $\widetilde{\br}_{l,1}$ (i.e., $\widetilde{r}_{l,0,k}$ and $\widetilde{r}_{l,1,k}$) obtained when channel is sensed as idle and busy, respectively.

From the two separate minimization problems in (\ref{mmse_single_pilot2}), it is well-known that the optimal MMSE estimates are given by the conditional expectations of $\br_l$ given the observation $Y_{l}$ and the sensing decision $\widehat{\mathcal{H}}_j$ for $j=0,1$, i.e., we have
\begin{align}
\widehat{\br}_{l,j,\text{mmse}}=\mathbb{E}\{\br_l|Y_{l},\widehat{\mathcal{H}}_{j}\}.
\end{align}
The MMSE estimates can further be formulated in terms of the conditional expectations of $\br_l$ given the true hypothesis $\mathcal{H}_{0}$ or $\mathcal{H}_{1}$, sensing decision $\widehat{\mathcal{H}}_j$ and the observation $Y_{l}$ as follows:
\begin{align}\label{estimate_single_pilot}
\widehat{\br}_{l,j,\text{mmse}} &= \mathbb{E}\{\br_l|Y_{l},\widehat{\mathcal{H}}_{j}\}
=\Pr\{\mathcal{H}_{0}|\widehat{\mathcal{H}}_{j},Y_{l}\}\mathbb{E}\{\br_l|Y_{l},\mathcal{H}_{0},\widehat{\mathcal{H}}_{j}\}+\Pr\left\{\mathcal{H}_{1}|\widehat{\mathcal{H}}_{j},Y_{l}\right\}\mathbb{E}\{\br_l|Y_{l},\mathcal{H}_{1},\widehat{\mathcal{H}}_{j}\}.
\end{align}
From Bayes' rule, we can express the above conditional probabilities as
\begin{align}\label{P1}
\Pr\{\mathcal{H}_{0}|\widehat{\mathcal{H}}_{j},Y_{l}\}&=\frac{\Pr\{\mathcal{H}_{0}\}\Pr \{\widehat{\mathcal{H}}_{j}|\mathcal{H}_{0}\}f(Y_{l}|\widehat{\mathcal{H}}_{j},\mathcal{H}_{0})}{\sum_{i=0}^{1}\Pr \{\mathcal{H}_{i}\}\Pr\{\widehat{\mathcal{H}}_{j}|\mathcal{H}_{i}\}f(Y_{l}|\widehat{\mathcal{H}}_{j},\mathcal{H}_{i})}=1-\Pr\{\mathcal{H}_{1}|\widehat{\mathcal{H}}_{j},Y_{l}\}.
\end{align}
Note that $\Pr\{a|b,c\}$ above denotes the conditional probability of $a$ given $b$ and $c$, and $f(a|b,c)$ is the conditional distribution of $a$ given $b$ and $c$.

If the fading coefficients, background noise, and primary users' faded received signal are all Gaussian distributed as we considered, the conditional expectations in (\ref{estimate_single_pilot})
are given by
\begin{align*}
\mathbb{E}\{\br_l|Y_{l},\mathcal{H}_{i},\widehat{\mathcal{H}}_{j}\}&=\mathbb{E}\{\br_lY_{l}^*|\mathcal{H}_{i},\widehat{\mathcal{H}}_{j}\}\mathbb{E}\{Y_{l}Y_{l}^{*}|\mathcal{H}_{i},\widehat{\mathcal{H}}_{j}\}^{-1}Y_{l}.
\end{align*}
For the special case where only the last pilot symbol is employed in channel estimation (i.e., $K=1$), we have
\begin{align}
\mathbb{E}\{\br_{l}|y_{lM}\mathcal{H}_{i},\widehat{\mathcal{H}}_{j}\}&=\frac{\mathbb{E}\{\br_{l}y_{lM}^*|\mathcal{H}_{i},\widehat{\mathcal{H}}_{j}\}} {\mathbb{E}\{|y_{lM}|^{2}|\mathcal{H}_{i},\widehat{\mathcal{H}}_{j}\}}y_{lM}
\nonumber\\
&=\frac{\mathbb{E}\{\br_{l}[\q_{l,t,j}\br_{l}+w_{lm}]^*|\mathcal{H}_{i},\widehat{\mathcal{H}}_{j}\}}{\mathbb{E}\{|\q_{l,t,j}\br_{l}+w_{lm}|^{2}|\mathcal{H}_{i},\widehat{\mathcal{H}}_{j}\}}y_{lM}\nonumber\\
&=\frac{\mathbb{E}\{\br_{l}\br_{l}^{\dagger}\q_{l,t,j}^{\dagger}|\mathcal{H}_{i},\widehat{\mathcal{H}}_{j}\}}{\mathbb{E}\{\q_{l,t,j}\br_{l}\br_{l}^{\dagger}\q_{l,t,j}^{\dagger}+w_{lm}w_{lm}^{*}|\mathcal{H}_{i},\widehat{\mathcal{H}}_{j}\}}y_{lM}\nonumber\\
&=\frac{\mathbb{E}\{\br_{l}\br_{l}^{\dagger}\}\q_{l,t,j}^{\dagger}}{\q_{l,t,j}\mathbb{E}\{\br_{l}\br_{l}^{\dagger}\}\q_{l,t,j}^{\dagger}+\sigma_{w,i}^{2}}y_{lM}\nonumber\\
&=\frac{\Lambda_{\br_{l}}\q_{l,t,j}^\dagger}{\mathcal{E}_{t,j}\sigma_r^2+\sigma_{w,i}^2}y_{lM},
\end{align}
where
\begin{equation}
\sigma_{w,i}^2=\left\{
  \begin{array}{ll}
    \sigma_n^2, & i=0 \\
    \sigma_n^2+\sigma_s^2, & i=1
  \end{array}
\right.,
\end{equation}
$\Lambda_{\br_{l}}$ is the covariance matrix of $\br_{l}$, and the superscripts $*$ and $\dagger$ are used to denote the conjugate and conjugate transpose, respectively.
\subsection{L-MMSE Estimation}
As seen in (\ref{estimate_single_pilot}), the MMSE estimates are given in terms of the weighted combination of the conditional expectations of the fading vector $\br_{l}$ given the observation, sensing decision, and the true state of the primary user activity. These conditional expectations in general do not have closed-form expressions unless Gaussian fading and Gaussian noise are considered. Additionally, even in a Gaussian setting, evaluation of the mean-squared error of the MMSE estimation is a difficult task due to the presence of the conditional distributions of $Y_{l}$ in (\ref{P1}). More specifically, while the conditional expectations $\mathbb{E}\{\br_{l}|\mathcal{H}_{i},\widehat{\mathcal{H}}_{j},Y_{l}\}$ for $i,j \in \{0,1\}$ are linear functions of the observation $Y_{l}$, the MMSE estimates $\widehat{\br}_{l,j,\text{mmse}}$ for $j = 0,1$ are complicated non-linear functions of $Y_{l}$ due to having $f(Y_{l}|\widehat{\mathcal{H}}_{j},\mathcal{H}_{i})$ in the conditional probabilities in (\ref{P1}).

Motivated by these considerations, we now study linear MMSE (L-MMSE) estimation as a suboptimal method that can alleviate the concerns regarding the complexity of the optimal MMSE estimation but can still perform close to being optimal. As also briefly discussed above, it is important to note that MMSE and L-MMSE estimators are in general different in our cognitive radio setting even if fading $\br_{l}$, noise $N_{l}$, and the primary users' received faded signal $S_{l}$ are all Gaussian distributed. The primary reason for this is that the experienced additive disturbance, which is either $W_{l} = N_{l}+S_{l}$ or $W_{l} = N_{l}$, has a Gaussian mixture distribution in the presence of sensing uncertainty.

The L-MMSE estimate when the channel sensing result is $\widehat{\mathcal{H}}_{j}$ is
\begin{align}\label{lMMSE_single_H0_Estimation}
\widehat{\br}_{l,j, \text{lmmse}}&=\Lambda_{\br_{l}Y_{l}|\widehat{\mathcal{H}}_{j}}\Lambda_{Y_{l}|\widehat{\mathcal{H}}_{j}}^{-1}Y_{l}=\Lambda_{\br_{l}}\q_{l,t,j}^{*}\left[\sum_{i=0}^{1}\Pr\{\mathcal{H}_{i}|\widehat{\mathcal{H}}_{j}\} \Lambda_{Y_{l}|\widehat{\mathcal{H}}_{j},\mathcal{H}_{i}}\right]^{-1}Y_{l}=\Lambda_{l,j}Y_{l}.
\end{align}
where $\Lambda_{\br_{l}Y_{l}|\widehat{\mathcal{H}}_{i}}$ and $\Lambda_{Y_{l}|\widehat{\mathcal{H}}_{i},\mathcal{H}_j}$ are the cross-correlation matrix of $\br_{l}$ and $Y_{l}$, and the variance of $Y_{l}$, respectively, given that the channel sensing result is $\widehat{\mathcal{H}}_{j}$ and the true hypothesis is $\mathcal{H}_i$. Now, the MSE of L-MMSE estimation is given by
\begin{align}\label{mse_single_pilot_lmmse}
\text{MSE}=&\sum_{j=0}^{1}\Pr\{\widehat{\mathcal{H}}_{j}\}\mathbb{E}\{\|\underbrace{\br_{l}-\widehat{\br}_{l,j}}_{\widetilde{\br}_{l,j}}\|^{2}|\widehat{\mathcal{H}}_{j}\}=\sum_{j=0}^{1}\sum_{i=0}^{1}\Pr\{\mathcal{H}_{i}\}\Pr\{\widehat{\mathcal{H}}_{j}|\mathcal{H}_{i}\}\mathbb{E}\{\|\widetilde{\br}_{l,j}\|^{2}|\widehat{\mathcal{H}}_{j},\mathcal{H}_{i}\}
\end{align}
where, for the special case when only the last pilot symbol is employed in channel estimation (i.e., $K=1$), we have
\begin{align*}
\mathbb{E}\{\|\widetilde{\br}_{l,j}\|^{2}|&\widehat{\mathcal{H}}_{j},\mathcal{H}_{i}\}= M\sigma_{r}^{2}-2\Lambda^{*}_{l,j}\Lambda_{\br_{l}}\q_{l,t,j}+\Lambda_{l,j}^{*}\Lambda_{l,j}\mathbb{E}\{|y_{lM}|^{2} |\widehat{\mathcal{H}}_{j},\mathcal{H}_{i}\}.
\end{align*}

\section{Achievable Rates, Power Allocation, and Training Period}\label{achievable_rates}
Having addressed channel sensing and channel estimation, we next study the achievable rates in the cognitive radio channel when the secondary users are equipped with only imperfect channel sensing and estimation results. We note that an intricate relationship exists between sensing reliability, channel estimation quality, and transmission rates. In general, sensing performance affects the structure and quality of the channel estimation, which in turn have implications on the achievable rates.

\subsection{Achievable Rates with Linear Modulation Schemes}\label{achievable_rates_bpsk}
In this subsection, we consider that memoryless linear modulation schemes (e.g., pulse amplitude modulation (PAM), phase-shift keying (PSK) or quadrature amplitude modulation (QAM)) are employed at the secondary transmitter to send information to the secondary receiver in the data transmission phase with imperfect CSI obtained by performing single MMSE estimation at the secondary receiver. We assume that CSI is available only at the secondary receiver. When the channel is sensed as idle (i.e., the sensing decision is $\widehat{\mathcal{H}}_{0}$), the secondary transmitter sends the signal $x_{0,u,k}$ in the $k^{\text{th}}$ symbol interval where $Ml+1\leq k\leq M(l+1)-1$, $ 1 \le u \le U$, and $U$ is the modulation size. Under sensing decision $\widehat{\mathcal{H}}_{0}$, the average energy of the modulation is $\mathcal{E}_{d,0} = \sum_{u = 1}^U p_u |x_{0,u}|^2$ where $p_u$ denotes the prior probability of $x_{0,u}$. On the other hand, when the channel sensing decision is $\widehat{\mathcal{H}}_{1}$, the secondary transmitter sends the signal $x_{1,u, k}$ with average energy $\mathcal{E}_{d,1} = \sum_{u = 1}^U p_u |x_{1,u}|^2$. Note that the data transmission is performed after the channel is estimated in the training phase.

Over a duration of $M$ symbols, an achievable rate expression can be obtained by considering the mutual information between the input and output vectors:
\begin{align}
\frac{1}{M}\mathbb{E}\left[I\{\mathbf{x}_{lM};\mathbf{y}_{lM}|\widehat{\br}_{lM}\}\right]&\geq\frac{1}{M}\sum_{k=lM+1}^{lM+M-1}\mathbb{E}\left[I(x_{k};y_{k}|\hat{r}_{k})\right] \label{Capacity_Lower_Bound_1}
\\
&=\frac{1}{M}\sum_{k=lM+1}^{lM+M-1}\sum_{j=0}^{1}\mathbb{E}\bigg[\Pr\{\widehat{H}_{j}\}I(x_{j,u,k};y_{k}|\hat{r}_{j,k},\widehat{H}_{j})\bigg]\label{Capacity_Lower_Bound_8}
\end{align}
On the left-hand side of (\ref{Capacity_Lower_Bound_1}), $\mathbf{x}_{lM} = [x_{lM+1}, x_{lM+2}, \ldots, x_{lM+M-1}]$ and $\mathbf{y}_{lM}$ denote the $M-1$ (or equivalently $TB-1$) dimensional input data and output vectors, respectively, in each transmission block of $T$ seconds and the expectation is with respect to $\widehat{\br}_{lM}$, which is the vector of the estimates of corresponding channel fading coefficients. The lower bound in (\ref{Capacity_Lower_Bound_1}) is due to the fact that a channel with memory has a higher reliable communication rate than the memoryless channel with the same marginal transition probability \cite{jun_chen}. Hence, the mutual information between the input and output vectors is in general larger than the sum of the symbol-wise mutual information terms given on the right-hand side of (\ref{Capacity_Lower_Bound_1}). Finally, (\ref{Capacity_Lower_Bound_8}) is obtained by conditioning the mutual information on the sensing decisions. Henceforth, our achievable rate analysis is based on the achievable rate expression in (\ref{Capacity_Lower_Bound_8}).

For a linear modulation scheme with $U$ signals, the input-output mutual information given the channel estimate and the channel sensing decision $\widehat{\mathcal{H}}_{j}$ can be expressed as
\begin{align}
&I(x_{j,u,k};y_{k}|\hat{r}_{j,k},\widehat{\mathcal{H}}_{j})=\sum_{u = 1}^U p_u \int f(y_{k}|x_{j,u,k}, \hat{r}_{j,k}, \widehat{\mathcal{H}}_{j}) \log\frac{f(y_{k}|x_{j,u,k}, \hat{r}_{j,k}, \widehat{\mathcal{H}}_{j})}{f(y_{k}|\hat{r}_{j,k}, \widehat{\mathcal{H}}_{j})} \, d y_k \nonumber
\end{align}
where
\begin{align*}
f(y_{k}|x_{j,u,k}, \hat{r}_{j,k}, \widehat{\mathcal{H}}_{j})=&\Pr\{\mathcal{H}_{0}|\widehat{\mathcal{H}}_{j}\}f(y_{k}|x_{j,u,k}, \hat{r}_{j,k}, \widehat{\mathcal{H}}_{j},\mathcal{H}_{0})+\Pr\{\mathcal{H}_{1}|\widehat{\mathcal{H}}_{j}\}f(y_{k}|x_{j,u,k}, \hat{r}_{j,k}, \widehat{\mathcal{H}}_{j},\mathcal{H}_{1})
\end{align*}
and
\begin{equation*}
f(y_{k}|x_{j,u, k}, \hat{r}_{j,k}, \widehat{\mathcal{H}}_{j},\mathcal{H}_{i})=\frac{1}{\pi\sigma_{j,k,i}^2}\exp\left(-\frac{|y_{k}-\hat{r}_{j,k}x_{j,u,k}|^2}{\pi\sigma_{j,k,i}^2}\right)
\end{equation*}
with
\begin{equation*}
\sigma_{j,k,i}^2=\left\{
  \begin{array}{ll}
    \sigma_n^2+\sigma_{\tilde{r}_{j,k}}^2|x_{j,u,k}|^2, & i=0, \\
    \sigma_n^2+\sigma_s^2+\sigma_{\tilde{r}_{j,k}}^2|x_{j,u,k}|^2, & i=1.
  \end{array}
\right.
\end{equation*}
We know that the total average energy in one training and data transmission block is $M\overline{P}_0/B$ or $M\overline{P}_1/B$ (or equivalently $\overline{P}_0 T$ or $\overline{P}_1 T$ recalling that $M = TB$) when the channel is sensed as idle or busy, respectively, over an interval of $M$ symbols. If $\mu_i$ fraction of the total energy is allocated to the training symbol, we have the pilot energy given by
\begin{equation}\label{training_power}
\mathcal{E}_{t,i}=\frac{\mu_iM\bar{P}_i}{B}
\end{equation}
for $i=0,1$. Note that $\mu_0$ and $\mu_1$ are fractions when the channel sensing results are $\widehat{\mathcal{H}}_0$ and $\widehat{\mathcal{H}}_1$, respectively. Remaining energy is assumed to be equally allocated among the data symbols. Hence, energy per data symbol is
\begin{equation}\label{data_power}
\mathcal{E}_{d,i}=\frac{(1-\mu_i)M\bar{P}_i}{B(M-1)}
\end{equation}
for $i=0,1$. In (\ref{training_power}) and (\ref{data_power}), $M$, $\mu_{0}$, and $\mu_{1}$ are the design parameters that control the pilot assisted cognitive radio transmissions. $M$ is the frequency of the pilot symbol, and $\mu_0$ and $\mu_1$ are the fractions of the total energy, dedicated to channel training when the channel sensing decisions are $\widehat{\mathcal{H}}_0$ and $\widehat{\mathcal{H}}_1$, respectively. Clearly, the training parameters have an impact on the quality of data transmissions. For instance, improving the estimation quality with increasingly more frequent pilot symbol transmissions or by allocating more energy to training eventually reduces the data transmission rate since a smaller duration or smaller energy is allocated for data transmission. On the other hand, decreasing the pilot frequency or the training energy beyond a threshold decreases the data transmission rate as well due to the degradation of the channel estimate quality.

For given $\overline{P}_0$, $\overline{P}_1$, and channel fading statistics, the training period, $M$, and training energy fractions, $\mu_0$, $\mu_1$, which maximize the achievable rates, can be determined by solving the optimization problem given in
\begin{align}\label{eq:optimizationproblem}
(\mu_0^*,\mu_1^*,M^*)
=&\arg\max_{1 \leq M \leq (Q-N)B}\sum_{j=0}^{1}\Pr\{\widehat{\mathcal{H}}_j\}\max_{0\leq\mu_j\leq1}\frac{1}{M}\sum_{k=lM+1}^{(l+1)M-1}\mathbb{E}\{I(x_{j,k};y_{k}|\hat{r}_{j,k},\widehat{\mathcal{H}}_j)\}.
\end{align}
Since closed-form solutions of this optimization problem are unlikely to be found, we resort to numerical techniques in Section \ref{numerical_results} to obtain the achievable-rate maximizing training and transmission parameters. For ease of exposition and computation, in the numerical results, we consider equiprobable binary PSK (BPSK) modulation with signals $x_{0,1,k}=-\sqrt{\mathcal{E}_{d,0}}$ and $x_{0,2,k}=\sqrt{\mathcal{E}_{d,0}}$ under $\widehat{\mathcal{H}}_{0}$ and signals $x_{1,1,k}=-\sqrt{\mathcal{E}_{d,1}}$ and $x_{1,2,k}=\sqrt{\mathcal{E}_{d,1}}$ under $\widehat{\mathcal{H}}_{1}$.

\subsection{Achievable Rates with Gaussian Signaling}\label{achievable_rates_gaussian}
In this subsection, we assume that the channel input signals are Gaussian distributed. As discussed before, in the presence of sensing errors, the additive disturbance has a Gaussian mixture nature. For instance, given that the channel sensing result is $\widehat{\mathcal{H}}_j$ and the input is $x_{j,k}$, the additive disturbance is either $z_{j,k}=\tilde{r}_{j,k}x_{j,k}+n_{k}$ or $z_{j,k}=\tilde{r}_{j,k}x_{j,k}+n_{k}+s_{k}$ depending on whether the channel is actually idle or busy, respectively, and it has the following Gaussian mixture conditional probability density function:
\begin{align}
f(z_{j,k}|\widehat{\mathcal{H}}_j, x_{j,k})=&\frac{\Pr\{\mathcal{H}_0|\widehat{\mathcal{H}}_j\}}{\pi(\sigma_{\tilde{r}_{j,k}}^2|x_{j,k}|^2+\sigma_n^2)} \exp\left(-\frac{|z_{j,k}|^2}{\sigma_{\tilde{r}_{j,k}}^2|x_{j,k}|^2+\sigma_n^2}\right) \nonumber
\\
&+\frac{\Pr\{\mathcal{H}_1|\widehat{\mathcal{H}}_j\}}{\pi(\sigma_{\tilde{r}_{j,k}}^2|x_{j,k}|^2+\sigma_n^2+\sigma_s^2)} \exp\left(-\frac{|z_{j,k}|^2}{\sigma_{\tilde{r}_{j,k}}^2|x_{j,k}|^2+\sigma_n^2+\sigma_s^2}\right). \label{w_0}
\end{align}
Note that we have included the channel estimate error in the additive disturbance above. Now, for any given input distribution, the input-output mutual information can be expressed as
\begin{align}\label{mutual_information_0}
I(x_k;y_k|\hat{r}_k,\widehat{\mathcal{H}}_j)&=h(y_k|\hat{r}_k,\widehat{\mathcal{H}}_j)-h(y_k|x_k,\hat{r}_k,\widehat{\mathcal{H}}_j)\nonumber\\
&=h(y_k|\hat{r}_k,\widehat{\mathcal{H}}_j)-h(z_k|x_k,\widehat{\mathcal{H}}_j)
\end{align}
where $h(\cdot)$ is the differential entropy. Note that for the Gaussian-mixture distributed $z_{k}$ above, we need to evaluate the differential entropy, which does not admit a closed-form expression. Hence, achievable rate and capacity expressions are not readily available. However, in the following result, we employ several bounding techniques and derive a closed-form achievable rate expression when the input is Gaussian distributed. This expression explicitly depends on the channel estimate, variance of the estimation error, transmission energies, and the channel sensing reliability via the probabilities $\Pr\{\widehat{\mathcal{H}}_j\}$ and $\Pr\{\mathcal{H}_i|\widehat{\mathcal{H}}_{j}\}$.

\begin{theo} \label{lower_bound_achievable_rate}
For the cognitive radio channel with channel sensing and channel estimation errors, an achievable rate expression is given by
\begin{align}\label{capacity_lower_bound}
R(M^*,\mu_0^*,\mu_1^*)=\max_{1 \leq M \leq (Q-N)B}&\sum_{j=0}^{1}\Pr\{\widehat{\mathcal{H}}_j\}\max_{0\leq\mu_j\leq1}\frac{1}{M}\nonumber\\&\times\sum_{k=lM+1}^{(l+1)M-1}\mathbb{E}\left\{\log\left(1+ \frac{|\hat{r}_{j,k}|^2\mathcal{E}_{d,j}}{\sigma_{\tilde{r}_{j,k}}^2\mathcal{E}_{d,j}+\sigma_{n}^2+\Pr\{\mathcal{H}_1|\widehat{\mathcal{H}}_{j}\}\sigma_s^2}\right)\right\}
\end{align}
where $\mathcal{E}_{d,j}=\frac{(1-\mu_j)M\bar{P}_j}{B(M-1)}$ is the data symbol energy.
\end{theo}

\emph{Proof:} We start with the sum of symbol-wise mutual information expressions over a duration of $M$ symbols:
\begin{align}\label{lower_bound_proof}
\frac{1}{M}\sum_{j=0}^{1}\sum_{k=lM+1}^{(l+1)M-1}\mathbb{E}\bigg[&\Pr\{\widehat{\mathcal{H}}_j\}I(x_{j,k};y_{k}|\hat{r}_{j,k},\widehat{\mathcal{H}}_j)\bigg].
\end{align}
In order to establish lower bounds on the conditional mutual information expressions in (\ref{lower_bound_proof}), we follow the approach used in \cite{medard}. Now, let us consider the mutual information when the channel sensing result is $\widehat{\mathcal{H}}_j$, and express it in terms of differential entropies:
\begin{align}
I(x_{j,k};y_{k}|\hat{r}_{j,k},\widehat{\mathcal{H}}_{j})&=h(x_{j,k}|\hat{r}_{j,k},\widehat{\mathcal{H}}_{j})-h(x_{j,k}|y_{k},\hat{r}_{j,k},\widehat{\mathcal{H}}_{j})\\
&=h(x_{j,k}|\widehat{\mathcal{H}}_{j})-h(x_{j,k}|y_{k},\hat{r}_{j,k},\widehat{\mathcal{H}}_{j}). \label{I_0_proof}
\end{align}
An upper bound on $h(x_{j,k}|y_{k},\hat{r}_{j,k},\widehat{\mathcal{H}}_{j})$ can be found as
\begin{align}
h(x_{j,k}|y_{k},\hat{r}_{j,k},\widehat{\mathcal{H}}_{j})&=h(x_{j,k}-\beta y_{k}|y_{k}, \hat{r}_{j,k},\widehat{\mathcal{H}}_{j})\label{bound_h_0}\\
&\leq h(x_{j,k}-\beta y_{k} | \hat{r}_{j,k},\widehat{\mathcal{H}}_{j})\label{bound_h_1}\\
&\leq\log\left(\pi e \, \text{var}(x_{j,k}-\beta y_{k}| \hat{r}_{j,k},\widehat{\mathcal{H}}_{j})\right)\label{bound_h_2}
\end{align}
for any $\beta$. (\ref{bound_h_0}) is due to the fact that adding a constant does not affect the entropy. Since conditioning always decreases the entropy, we have (\ref{bound_h_1}). We know that the entropy of a random variable with given variance is upper-bounded by the entropy of a Gaussian random variable with the same variance, and hence we obtain (\ref{bound_h_2}). In order tighten the bound in (\ref{bound_h_2}), we minimize $\text{var}(x_{j,k}-\beta y_{k}| \hat{r}_{j,k},\widehat{\mathcal{H}}_{j})$ over $\beta$. With this purpose, we pick $\beta$ such that $\beta y_{k}$ is the L-MMSE estimate of $x_{j,k}$ in terms of $y_{k}$, which yields
\begin{align*}
&\text{var}(x_{j,k}-\beta y_{k} | \hat{r}_{j,k},\widehat{\mathcal{H}}_{j})=\frac{\sigma_{\tilde{r}_{j,k}}^2\mathcal{E}_{d,j}^2+[\sigma_{n}^{2}+\Pr\{\mathcal{H}_1|\widehat{\mathcal{H}}_j\}\sigma_s^2]\mathcal{E}_{d,j}}
{|\hat{r}_{j,k}|^2\mathcal{E}_{d,j}+\sigma_{\tilde{r}_{j,k}}^2\mathcal{E}_{d,j}+\sigma_{n}^{2}+\Pr\{\mathcal{H}_1|\widehat{\mathcal{H}}_j\}\sigma_s^2}.
\end{align*}
Hence, we have
\begin{align}
&h(x_{j,k}|y_{k},\hat{r}_{j,k},\widehat{\mathcal{H}}_{j})\leq\log\left(\pi e \frac{\sigma_{\tilde{r}_{j,k}}^2\mathcal{E}_{d,j}^2+[\sigma_{n}^{2}+\Pr\{\mathcal{H}_1|\widehat{\mathcal{H}}_j\}\sigma_s^2]\mathcal{E}_{d,j}}
{|\hat{r}_{j,k}|^2\mathcal{E}_{d,j}+\sigma_{\tilde{r}_{j,k}}^2\mathcal{E}_{d,j}+\sigma_{n}^{2}+\Pr\{\mathcal{H}_1|\widehat{\mathcal{H}}_j\}\sigma_s^2}\right).\label{bound_h_2_2}
\end{align}
Inserting the upper bound in (\ref{bound_h_2_2}) into (\ref{I_0_proof}) and noting that $x_{j,k}$ is assumed to be Gaussian distributed and hence $h(x_{j,k}|\widehat{\mathcal{H}}_{j}) = \log\left(\pi e \mathcal{E}_{d,j}\right)$, we can lower bound the mutual information between the input $x_{j,k}$ and the output $y_{k}$ as
\begin{align}
I(x_{j,k};y_{k}|\hat{r}_{j,k},\widehat{\mathcal{H}}_{j})&\geq\log\left(\pi e \mathcal{E}_{d,j}\right)-\log\left(\pi e \frac{\sigma_{\tilde{r}_{j,k}}^2\mathcal{E}_{d,j}^2+[\sigma_{n}^{2}+\Pr\{\mathcal{H}_1|\widehat{\mathcal{H}}_j\}\sigma_s^2]\mathcal{E}_{d,j}}
{|\hat{r}_{j,k}|^2\mathcal{E}_{d,j}+\sigma_{\tilde{r}_{j,k}}^2\mathcal{E}_{d,j}+\sigma_{n}^{2}+\Pr\{\mathcal{H}_1|\widehat{\mathcal{H}}_j\}\sigma_s^2}\right)\nonumber\\
&=\log\left(1+\frac{|\hat{r}_{j,k}|^2\mathcal{E}_{d,j}}{\sigma_{\tilde{r}_{j,k}}^2\mathcal{E}_{d,j} +\sigma_{n}^{2}+\Pr\{\mathcal{H}_1|\widehat{\mathcal{H}}_j\}\sigma_s^2}\right). \label{lower_bound_achievable_rate_I0}
\end{align}
Plugging (\ref{lower_bound_achievable_rate_I0}) into (\ref{lower_bound_proof}) and optimizing over the training parameters, we obtain the achievable rate expression in (\ref{capacity_lower_bound}). \hfill
$\square$

\section{Numerical Results}\label{numerical_results}
In this section, we present our numerical results. We consider a channel with fading that is modeled as a first-order Gauss-Markov process whose dynamics is described by
\begin{equation}\label{gauss_markov}
r_{k}=\alpha r_{k-1}+\zeta_k\quad0\leq\alpha\leq1\quad k=1,2,...,
\end{equation}
where $\{\zeta_k\}$ is a sequence of i.i.d. circular complex Gaussian variables with zero-mean and variance equal to $(1-\alpha^{2})\sigma_{r}^{2}$. In (\ref{gauss_markov}), $\alpha$ is a parameter that controls the variations between the consecutive channel fading coefficients. For example, if $\alpha=1$, channel is constant, whereas, when $\alpha=0$, coefficients are varying independently. For bandwidths in the 10 kHz range, and Doppler spreads of the order of 100 Hz, $\alpha$ will range between 0.9 and 0.99 \cite{abou_faycal}. The auto-correlation function of $r$ is
\begin{equation}\label{autocorrelation}
R_{r}(n)=\alpha^{n}\sigma_{r}^{2}.
\end{equation}

Furthermore, in our numerical computations, unless indicated otherwise, we consider the following parameter values. We assume that the channel is busy with probability $0.2$ (i.e. $\Pr\{\mathcal{H}_{1}\}=0.2$ and $\Pr\{\mathcal{H}_{0}\}=0.8$). It is further assumed that the noise variance is $\sigma_{n}^{2}=1$, and the average power of interference is $\sigma_{s}^2=1$. Different values of the the channel variation parameter are considered but, if not specified explicitly, it is set to $\alpha=0.95$. The average channel power is $\sigma_{r}^{2}=1$. We set the probability of detection $P_d = \Pr\{\widehat{\mathcal{H}}_{1}|\mathcal{H}_{1}\}=0.9$ and the probability of false alarm $P_f = \Pr\{\widehat{\mathcal{H}}_{1}|\mathcal{H}_{0}\}=0.2$. We consider the period of pilot symbols to be $M=10$. Moreover, when the channel is sensed as busy, the pilot power is set to $\mathcal{E}_{t,1}=1$, and when the channel is sensed as idle, the pilot power is set to $\mathcal{E}_{t,0}=10$. Hence, if the channel is sensed busy, the pilot power is dropped by $10$ dB. We finally note that we consider a single pilot-symbol estimation technique in our numerical results.

\begin{figure}
\begin{center}
\includegraphics[width =\figsize\textwidth]{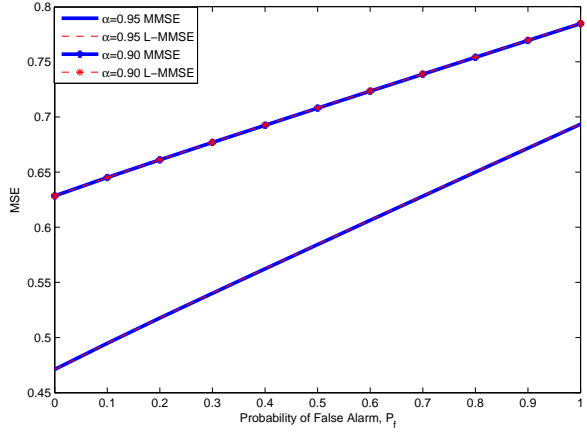}
\caption{MSE vs. Probability of False Alarm, $P_{f}$, when $P_{d}=0.9$, $\sigma_{s}^2=1$, $\sigma_n^2=1$, $\sigma_r^2=1$, $M=10$, $\mathcal{E}_{0}=10$, and $\mathcal{E}_{1}=1$.}\label{fig:fig1}
\end{center}
\end{figure}

\begin{figure}
\begin{center}
\includegraphics[width =\figsize\textwidth]{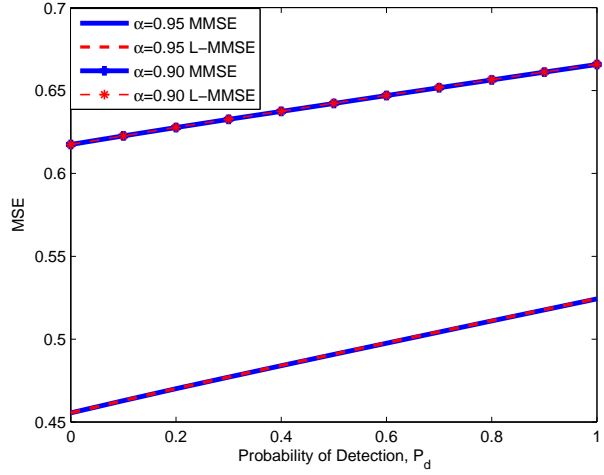}
\caption{MSE vs. Probability of Detection, $P_{d}$, when $P_{f}=0.2$, $\sigma_{s}^2=1$, $\sigma_n^2=1$, $\sigma_r^2=1$, $M=10$, $\mathcal{E}_{0}=10$, and $\mathcal{E}_{1}=1$.} \label{fig:fig2}
\end{center}
\end{figure}

In Fig. \ref{fig:fig1}, we plot the MSE in channel estimation as a function of probability of false alarm, $P_{f}$, for a fixed probability of detection, $P_{d}$ and for $\alpha = 0.90$ and $0.95$. We compare two different estimation methods. We observe that the MMSE and the L-MMSE methods give almost the same results, which is very promising, since L-MMSE estimation has a simpler structure. Furthermore, we see that the MSE is increasing with the increase in $P_{f}$. This is due to the fact that channel is sensed as busy more frequently due to increased probability of false alarms, and consequently the transmitter adjusts its pilot energy to the lower value, $\mathcal{E}_{t,1}$, more frequently with the goal of protecting the primary users. As a result, the estimation quality degrades. Note that since the primary users are not active in false alarm scenarios, the secondary users indeed miss the opportunity to transmit at high powers due to sensing uncertainty. We also notice that MSE expectedly increases as $\alpha$ drops from 0.95 to 0.90, resulting in a faster varying channel. This indeed is the common theme in all MSE curves. In Fig. \ref{fig:fig2}, we plot the MSE as a function of probability of detection, $P_{d}$, for a fixed probability of false alarm, $P_{f}$. Again, we see that L-MMSE and MMSE perform almost identically. Additionally, even though the channel sensing becomes more reliable with increasing $P_d$, MSE increases as well. This is again due to increased detection rates of primary user activity, leading the secondary transmitter to choose the lower energy, $\mathcal{E}_{t,1}$, more often. While this results in the degradation of the channel estimation quality, better protection of the primary users is achieved. Overall, both in Figs. \ref{fig:fig1} and \ref{fig:fig2}, we see that the channel estimation quality is critically dependent on the channel sensing performance.

\begin{figure}
\begin{center}
\includegraphics[width =\figsize\textwidth]{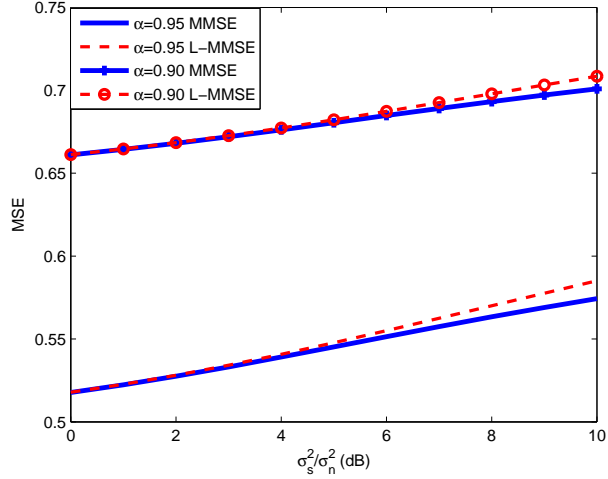}
\caption{MSE vs. $\sigma_s^2/\sigma_n^2$, when $P_{d}=0.9$, $P_{f}=0.2$, $M=10$, $\mathcal{E}_{0}=10$, and $\mathcal{E}_{1}=1$.} \label{fig:fig3}
\end{center}
\end{figure}

\begin{figure}
\begin{center}
\includegraphics[width =\figsize\textwidth]{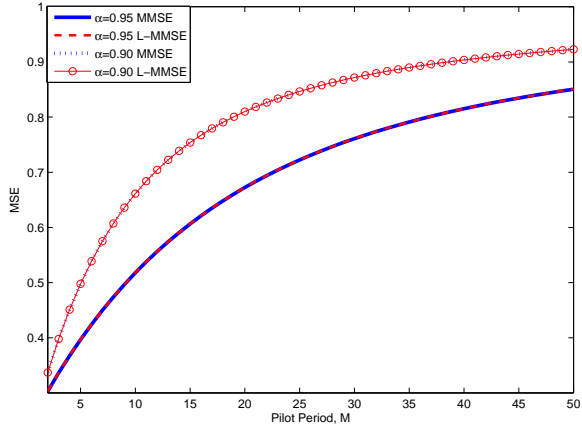}
\caption{MSE vs. Pilot Period, $M$, when $P_{d}=0.9$, $P_f=0.2$, $\sigma_{s}^2=1$, $\sigma_n^2=1$, $\sigma_r^2=1$, $\mathcal{E}_{0}=10$, $\mathcal{E}_{1}=1$.} \label{fig:fig4}
\end{center}
\end{figure}

In order to investigate the effects of the received signal interference from the primary users on the MSE, we plot the MSE vs. $\sigma_s^2/\sigma_n^2$ in Fig. \ref{fig:fig3}. We see that the performances of  MMSE and L-MMSE estimators are again close, with discrepancy increasing as $\sigma_s^2/\sigma_n^2$ becomes larger. Note that when $\sigma_s^2 = 0$, we have channel estimation in Gaussian noise in which case MMSE and L-MMSE estimators are identical. For $\sigma_s^2 > 0$, estimation is performed in Gaussian mixture noise due to sensing uncertainty. As $\sigma_s^2$ increases and $\sigma_s^2/\sigma_n^2$ grows, the departure from the Gaussian setting is emphasized and so is the difference between the performances of MMSE and L-MMSE estimators. In Fig. \ref{fig:fig3}, we also immediately notice that the MSE of both estimation schemes gets larger with increasing  $\sigma_s^2$. Obviously, in the presence of stronger interference from primary user transmissions, channel estimation is being performed in a noisier channel and MSE is higher.

In Fig. \ref{fig:fig4}, the MSEs of L-MMSE and MMSE estimations are plotted as a function of the pilot symbol period, $M$. It is clearly seen that the MSE is increasing with increasing $M$. Note that as $M$ increases, pilot symbols are sent less frequently and estimation quality for fading coefficients experienced long after the transmission of the pilot signal degrades severely. Indeed, the further away the channel fading coefficient is from the pilot symbol, with lesser quality the estimation is performed for that channel fading coefficient. Since we are considering the average MSE for each block, the average MSE increases with the increasing pilot symbol period.

\begin{figure}
\begin{center}
\includegraphics[width =\figsize\textwidth]{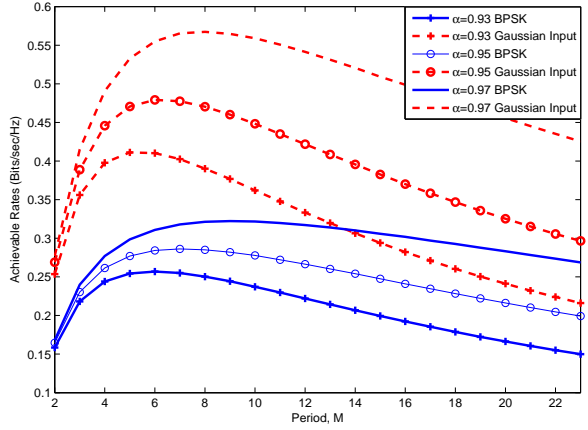}
\caption{Achievable Rates of BPSK and Gaussian inputs vs. training symbol period, $M$, when $P_{d}=0.9$, $P_f=0.2$, $\sigma_{s}^2=1$, $\sigma_n^2=1$, $\sigma_r^2=1$, $P_{0}=10$, $P_{1}=1$, $\mu_0=0.1$, and $\mu_1=0.1$.}\label{fig:fig5}
\end{center}
\end{figure}

\begin{figure}
\begin{center}
\includegraphics[width =\figsize\textwidth]{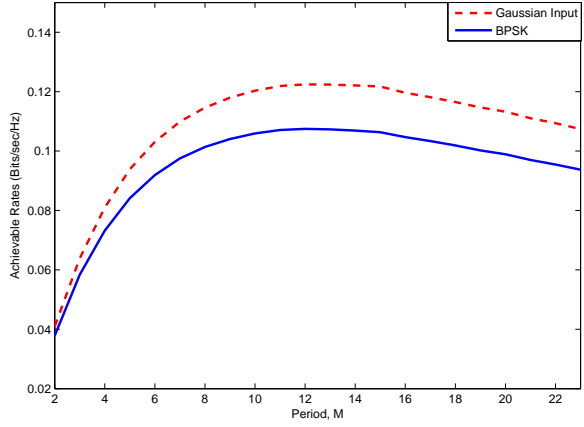}
\caption{Achievable Rates of BPSK and Gaussian inputs vs. training symbol period, $M$, when $P_{d}=0.9$, $P_f=0.2$, $\sigma_{s}^2=1$, $\sigma_n^2=1$, $\sigma_r^2=1$, $P_{0}=1$, $P_{1}=1$, $\mu_0=0.1$, $\mu_1=0.1$, and $\alpha = 0.95$.}\label{fig:fig6}
\end{center}
\end{figure}

\begin{figure}
\begin{center}
\includegraphics[width =\figsize\textwidth]{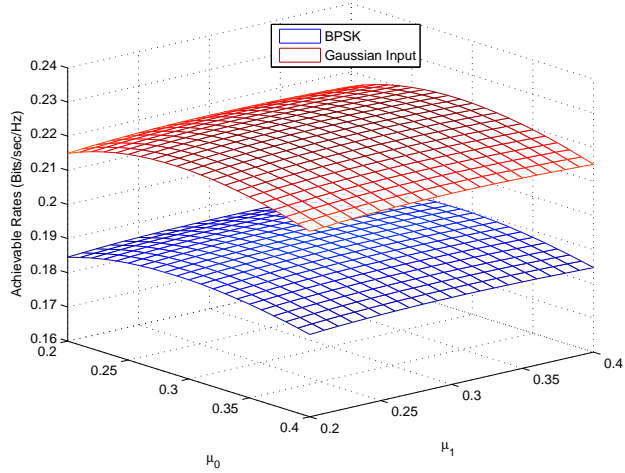}
\caption{Achievable Rates of BPSK and Gaussian inputs vs. training symbol ratios, $\mu_0$ and $\mu_1$, when $P_{d}=0.9$, $P_f=0.2$, $\sigma_{s}^2=1$, $\sigma_n^2=1$, $\sigma_r^2=1$, $P_{0}=10$, $P_{1}=1$, $M=12$ and $\alpha = 0.95$.}\label{fig:fig7}
\end{center}
\end{figure}

\begin{figure}
\begin{center}
\includegraphics[width =\figsize\textwidth]{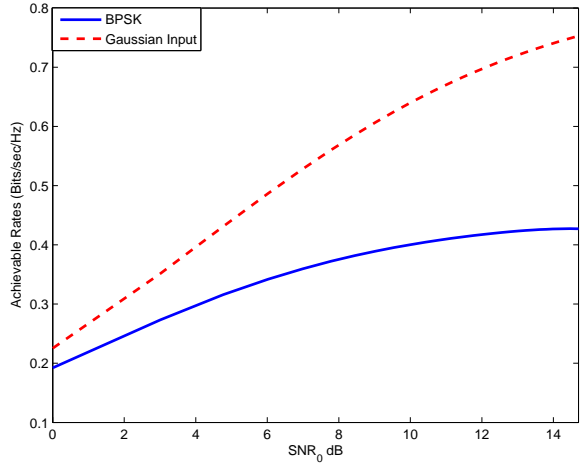}
\caption{Achievable Rates of BPSK and Gaussian inputs vs. $\tSNR_0=\frac{\overline{P}_0}{B\sigma_n^2}$, when $P_{d}=0.9$, $P_f=0.2$, $\sigma_{s}^2=1$, $\sigma_n^2=1$, $\sigma_r^2=1$, $P_{1}=1$, $M=12$, $\mu_0=0.29$, $\mu_1=0.31$, and $\alpha = 0.95$.}\label{fig:fig8}
\end{center}
\end{figure}

In Fig. \ref{fig:fig5}, we plot the achievable rates of BPSK and Gaussian input signaling vs. pilot symbol period, $M$ for $\alpha = 0.93, 0.95$, and $0.97$. Average input signal-to-noise ratio is $\tSNR_0=10\log_{10}\frac{\overline{P}_0}{B\sigma_n^2}=10\log_{10}(10) = 10$ dB when the channel is idle, and $\tSNR_1=10\log_{10}\frac{\overline{P}_1}{B(\sigma_n^2+\sigma_s^2)}=10\log_{10}(0.5)$ dB when the channel is busy. Note that the average input power is less when the channel is busy in order to protect the primary users. Furthermore, the fractions of training symbol energies are $\mu_0=\mu_1=0.1$. We observe that when $\alpha = 0.95$, the maximum achievable rates are obtained when the periods of pilot symbol transmissions are $M=7$ and $M=6$ for BPSK and Gaussian input, respectively, and the achievable rate of BPSK is almost half of the achievable rate of the Gaussian input. By employing modulations with larger constellations, the gap can be narrowed. We also note that when $\alpha$ decreases (e.g., $\alpha = 0.93$) and hence the channel varies faster, smaller achievable rates are attained and the rates are maximized at smaller values $M$, indicating that pilot symbols should be sent more frequently. In Fig. \ref{fig:fig6}, we set $\alpha = 0.95$ and plot the same rates when $\tSNR_0=10\log_{10}(1)$ dB while $\tSNR_1=10\log_{10}(0.5)$ dB. Hence, we have a lower value for $\tSNR_0$ now, which can arise due to more strict interference limitations. Rate-maximizing training pilot period for both types of input is now $M=12$. The gap between these two rates is less compared to that observed in Fig. \ref{fig:fig5}. In Fig. \ref{fig:fig7}, we plot the achievable rates as a function of $\mu_0$ and $\mu_1$. We observe that the maximum achievable rate with BPSK is obtained when $\mu_0=0.29$ and $\mu_1=0.31$. However, when the input is Gaussian, we observe that we attain the maximum achievable rate when $\mu_0=0.29$ and $\mu_1=0.30$. Finally, in Fig. \ref{fig:fig8}, the achievable rates of BPSK and Gaussian inputs maximized over $M$, $\mu_0$, and $\mu_1$ values are plotted as a function of $\tSNR_0$. As expected, achievable rates increase as $\tSNR_0$ increases. Also, while the achievable rate of the Gaussian input progressively grows, BPSK rate saturates due to the finite size of the constellation, and hence the gap between the maximum achievable rates is proportionally increasing with increasing $\tSNR_0$.

\begin{figure}
\begin{center}
\includegraphics[width =\figsize\textwidth]{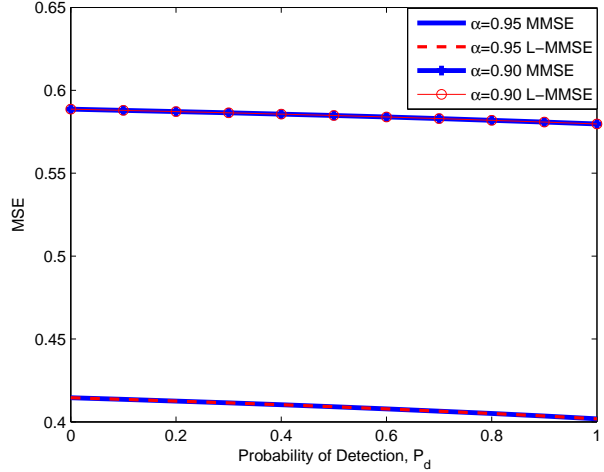}
\caption{MSE vs. Probability of Detection, $P_{d}$, when $P_{f}=0.2$, $\sigma_{s}^2=1$, $\sigma_n^2=1$, $\sigma_r^2=1$, $M=10$, and $\mathcal{E}_{0}=10$. Channel estimation is performed only when the channel is sensed as idle.} \label{fig:MSE_Idle}
\end{center}
\end{figure}

\begin{figure}
\begin{center}
\includegraphics[width =\figsize\textwidth]{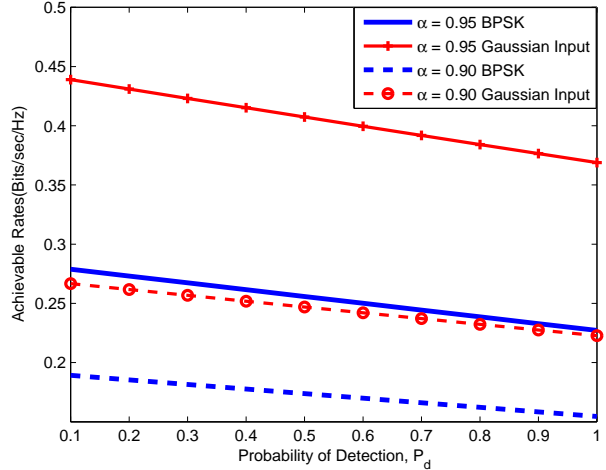}
\caption{Achievable Rates vs. Probability of Detection, $P_{d}$, when $P_{f}=0.2$, $\sigma_{s}^2=1$, $\sigma_n^2=1$, $\sigma_r^2=1$, $M=10$, and $P_{0}=10$. Secondary transmitter transmits only when the channel is sensed as idle.} \label{fig:Achievable_Idle}
\end{center}
\end{figure}

Heretofore in the numerical results, we have always considered the setting in which the secondary users communicate over both busy- and idle-sensed channels while keeping the transmission power lower in a busy-sensed channel. However, the analysis is easily applicable to an interweave scenario in which the secondary transmitter transmits only when the channel is sensed as idle. This can be accomplished by setting $\mathcal{E}_{t,1} = 0$ and $P_1 = 0$. In Figs. \ref{fig:MSE_Idle} and \ref{fig:Achievable_Idle}, we address this scenario. In Fig. \ref{fig:MSE_Idle}, we note that unlike in Fig. \ref{fig:fig2}, MSE decreases with increasing detection probability $P_d$. This is due to the following. In the case of miss-detections, secondary receiver performs channel estimation in the presence of interference from primary user transmissions and hence suffers from higher noise and experiences higher MSE with respect to that achieved when the channel is truly idle. As $P_d$ increases, miss-detection events occur less frequently and consequently channel estimation is affected less by the primary user interference. However, this does not necessarily improve the achievable rates. Indeed, as we observe in Fig. \ref{fig:Achievable_Idle}, intermittent transmission due to being silent in busy-sensed channels results in lower rates and the achievable rates diminish further with increasing $P_d$.

\section{Conclusion}\label{conclusion}

In this paper, we have considered mobile cognitive radio systems, operating over correlated fading channels with two different power levels in three phases, namely channel sensing, channel estimation, and data transmission. We have addressed a practical setting in which channel sensing and channel estimation are being performed with possible errors. We have initially addressed channel estimation in the presence of sensing uncertainty. We have derived the MMSE and L-MMSE estimators for a block of $M$ correlated fading coefficients and determined the MSE of the L-MMSE estimation. Through numerical analysis, we have demonstrated that the performance of L-MMSE estimator generally closely matches that of the MMSE estimator unless $\sigma_s^2/\sigma_n^2$ is large. Hence, due its simpler structure, we have noted that L-MMSE estimation can be preferred. Numerically, we have also investigated the impact of the channel sensing performance and training parameters on the MSE. For instance, increasing false-alarm or detection probabilities tend to increase the MSE of estimation if the secondary users communicate over both idle- and busy-sensed channels.  Furthermore, we have studied the achievable rates of linear modulation schemes and Gaussian inputs under both channel and sensing uncertainty. In particular, by using the sum of symbol-wise mutual information terms as our achievable rate expression, we have formulated the rates achieved by linear modulation schemes and also Gaussian input signals. We have derived a closed-form achievable rate expression for the Gaussian input in terms of the sensing reliability, channel estimate, variance of the channel estimation error, and data transmission energy. We have identified how achievable rates vary with the training parameters such as pilot period and pilot energies, and determined the rate-maximizing values of these parameters numerically.

\end{spacing}

\end{document}